\documentclass[12pt]{iopart}
\usepackage{iopams} 
\usepackage{setstack} 
\usepackage{latexsym}
\usepackage[dvipdfmx]{color,graphicx}

\newcommand{\eqb}{\begin{equation}}
\newcommand{\eqe}{\end{equation}}
\newcommand{\eqbnon}{\begin{equation*}}
\newcommand{\eqenon}{\end{equation*}}

\newcommand{\eqab}{\begin{eqnarray}}
\newcommand{\eqae}{\end{eqnarray}}
\newcommand{\eqabnon}{\begin{eqnarray*}}
\newcommand{\eqaenon}{\end{eqnarray*}}

\newcommand{\numb}{\numparts}   
\newcommand{\nume}{\endnumparts}

\newcommand{\defeq}{:=}
\newcommand{\defeqr}{=:}

\newcommand{\pd}[2]{\frac{\partial #1}{\partial #2}}
\newcommand{\od}[2]{\frac{{\rm d} #1}{{\rm d} #2}}
\newcommand{\diff}{{\rm d}}

\newcommand{\ls}[1]{_{\rm #1}}

\newcommand{\bound}{\mathcal{B}}

\begin{document}

\title{Maximum mass of a barotropic spherical star}

\author{Atsuhito Fujisawa}
\address{Department of Physics, Nagoya University, Chikusa-ku, 464-8602 Nagoya, Japan}
\ead{fujisawa@gravity.phys.nagoya-u.ac.jp}

\author{Hiromi Saida}
\address{Department of Physics, Daido University, Minami-ku, 457-8530 Nagoya, Japan}
\ead{saida@daido-it.ac.jp}

\author{Chul-Moon Yoo and Yasusada Nambu}
\address{Department of Physics, Nagoya University, Chikusa-ku, 464-8602 Nagoya, Japan}
\ead{yoo@gravity.phys.nagoya-u.ac.jp}
\ead{nambu@gravity.phys.nagoya-u.ac.jp}

\begin{abstract}
The ratio of total mass $m_\ast$ to surface radius $r_\ast$ of spherical perfect fluid ball has an upper bound, $G m_\ast/(c^2 r_\ast) \le \bound$. 
Buchdahl obtained $\bound\ls{Buch} = 4/9$ under the assumptions; non-increasing mass density in outward direction, and barotropic equation of state. 
Barraco and Hamity decreased the Buchdahl's bound to a lower value $\bound\ls{BaHa} = 3/8$ $(< 4/9)$ by adding the dominant energy condition to Buchdahl's assumptions. 
In this paper, we further decrease the Barraco-Hamity's bound to $\bound\ls{new} \simeq 0.3636403$ $(< 3/8)$ by adding the subluminal (slower-than-light) condition of sound speed. 
In our analysis, we solve numerically Tolman-Oppenheimer-Volkoff equations, and the mass-to-radius ratio is maximized by variation of mass, radius and pressure inside the fluid ball as functions of mass density.
\end{abstract}



\submitto{\CQG (accepted for publication)}

\maketitle

\section{Introduction and summary of result}
\label{sec:intro}

A question of self-gravitating compact object is the subject of this paper: 
How much weight or how small radius can a compact object possess? 
A quantitative answer is given by the mass-to-radius ratio $G m_\ast/(c^2 r_\ast)$, where $m_\ast$ and $r_\ast$ are respectively the total mass and surface radius of compact object. 
Assuming static spherical symmetry for simplicity, which implies the outside region of compact object is Schwarzschild geometry, this ratio is bounded above, $G m_\ast/(c^2 r_\ast) \le \bound$. 
The upper bound needs to satisfy, $\bound < 1/2$ $( \Leftrightarrow 2G m_\ast/c^2 < r_\ast )$, in order to avoid gravitational collapse.

An interesting issue of this bound is whether $\bound$ is less than $1/3$ or not. 
If an inequality, $\bound \ge 1/3$ $( \Leftrightarrow 3G m_\ast/c^2 \ge r_\ast )$, holds for static spherical case, unstable circular orbits of photons can appear in the outside Schwarzschild geometry. 
If such super-compact object, which possesses unstable circular orbits of photons but no black hole horizon, do neither emit nor reflect any radiation, then we cannot distinguish it from black holes by observing the so-called black hole shadows. 
Here, the black hole shadow is a dark region which is expected to appear in a fine image of optical/radio observation of black holes (see \cite{ref:takahashi.2004,ref:kanai+1.2013} and references therein): 
Consider a case that an optical source is extended behind a black hole and does not enter the inside of unstable circular orbits of photons. 
Observer cannot detect photons which passed through unstable circular orbits inward, because those photons are absorbed by black hole eventually. 
Hence, in the image of optical source, there should appear the dark region, on which those photons would be detected if black hole did not exist. 
This dark region is the black hole shadow, and the boundary of shadow is determined by the photons propagating on null geodesics winding many times around unstable circular orbits of photons. 
This means that a super-compact object possessing unstable circular orbits but no black hole horizon can provide us the same optical image of shadow with black holes. 
Since the resolution of image by radio observation is now approaching the visible angular size of largest black hole candidate~\cite{ref:miyoshi+4.2004,ref:doeleman+etal.2008}, an investigation of possible super-compact object seems to be an important issue for near future observational study of black holes. 
Therefore, an interesting and important issue is whether $\bound < 1/3$ holds or not.

Some exotic models of super-compact objects have been proposed such as gravastars, boson stars and so on. 
The gravastar as super-compact object has already been examined~\cite{ref:sakai+2.2014}, whereas the others remain to be examined. 
Those exotic models may be interesting. 
However, in this paper, we focus on a rather usual model.

Assuming that (i) the compact object is a static spherical ball of perfect fluid, (ii) its mass density is non-increasing with respect to radial coordinate, and (iii) its equation of state is barotropic, Buchdahl obtained $Gm_\ast/(c^2 r_\ast) \le \bound\ls{Buch} = 4/9$~\cite{ref:buchdahl.1959,ref:wald.1984}. 
Furthermore, by adding the dominant energy condition to Buchdahl's assumptions, Barraco and Hamity decreased the Buchdahl's bound to a lower value $\bound\ls{BaHa} = 3/8\, (< 4/9)$~\cite{ref:barraco+1.2002}. 
However, this upper bound is greater than $1/3$ $(< \bound\ls{BaHa})$.

Other works on the bound $\bound$ have been performed. 
For example, an effect of cosmological constant is examined~\cite{ref:boehmer.2005}, a case with infinite surface radius is analyzed~\cite{ref:hod.2007}, and a tangential pressure is considered~\cite{ref:andreasson.2008}. 
However, the Barraco-Hamity's bound, $\bound\ls{BaHa} = 3/8$, has not been lowered so far.

As explained in detail in section~\ref{sec:ratio}, we add the subluminal (slower-than-light) condition of sound speed to the assumptions of Buchdahl and Barraco-Hamity. 
We discuss, in section~\ref{sec:ratio}, how the subluminal-sound-speed condition restricts the form of equation of state, and also that this condition is more restrictive than the dominant energy condition. 
Then, our upper bound is lower than the Barraco-Hamity's bound,
\eqb
\label{eq:intro.result}
 \frac{G m_\ast}{c^2 r_\ast} \le \bound\ls{new} \simeq 0.3636403 \quad \Bigl( < \bound\ls{BaHa} = \frac{3}{8} \Bigr) \,.
\eqe
The reason why $\bound\ls{new}$ is approximate value is that we have performed numerical integration of Tolman-Oppenheimer-Volkoff equations. 
Although the upper bound is lowered, it still remains greater than $1/3$ $(< \bound\ls{new})$. 
Therefore, at present, we cannot deny the possibility that the same shadow image can be obtained from black holes and some super-compact object made of fluid matter. 
In order to sharpen the upper bound of $G m_\ast/(c^2 r_\ast)$, some other condition should be assigned to the form of equation of state. 
(We will report the case of polytropic equation of state in the other paper.)

Section~\ref{sec:ratio} is devoted to the details of our analysis. 
Section~\ref{sec:cd} is for conclusion and discussions.

\section{Variational analysis of mass-to-radius ratio}
\label{sec:ratio}

\subsection{Definitions and assumptions}
\label{sec:ratio.definition}

As explained in section~\ref{sec:intro}, we are interested in a static and spherically symmetric perfect fluid ball. 
A line element of spacetime is
\eqb
\label{eq:ratio.metric}
\fl
 \diff s^2 = g_{\mu\nu}\diff x^{\mu}\, \diff x^{\nu} =
 - e^{2\Phi(r)}c^2 \diff t^2 + \frac{\diff r^2}{1-2Gm(r)/(c^2 r)}
 + r^2 ( \diff \theta^2 + \sin^2\theta \diff \varphi^2 ) \,,
\eqe
where $(t,r,\theta,\varphi)$ is spherical poler coordinates, $\Phi(r)$ gives a lapse function, and $m(r)$ is a mass of perfect fluid contained in spherical region of radius $r$. 
The stress-energy-momentum tensor of perfect fluid is $T_{\mu\nu} = \sigma(r)c^2 u_\mu u_\nu + p(r)\, ( g_{\mu\nu} + u_\mu u_\nu )$, where $u = e^{-\Phi}\partial_{ct}$ is a four-velocity of static perfect fluid, and $\sigma(r)$ and $p(r)$ are respectively a mass density and pressure of perfect fluid.

By the regularity of spacetime at centre, a condition $m(0) = 0$ holds. 
This implies that the mass density at centre $\sigma\ls{c} = \sigma(0)$ is finite. 
We normalize all quantities by $\sigma\ls{c}$,
\eqb
\fl
\label{eq:ratio.normalization}
 R \defeq \frac{\sqrt{G \sigma\ls{c}}}{c}\,r
 \quad,\quad
 \Sigma(R) \defeq \frac{\sigma(r)}{\sigma\ls{c}}
 \quad,\quad
 M(R) \defeq \frac{\sqrt{G^3 \sigma\ls{c}}}{c^3}\,m(r)
 \quad,\quad
 P(R) \defeq \frac{p(r)}{\sigma\ls{c} c^2} \,.
\eqe
These are dimension-less. 
The lapse function, $\Phi(r) \defeq \Phi(R)$, does not need normalization because $\Phi$ is originally dimension-less by definition~\eref{eq:ratio.metric}.

Following Buchdahl~\cite{ref:buchdahl.1959,ref:wald.1984}, we adopt two assumptions. 
One of them is a barotropic equation of state,
\eqb
\label{eq:ratio.barotropy}
 P = P(\Sigma) \,.
\eqe
Hereafter, we regard the mass density $\Sigma$ as an independent variable, and the others are functions of it,
\eqb
 R=R(\Sigma) \quad,\quad M=M(\Sigma) \quad,\quad P=P(\Sigma) \quad,\quad \Phi=\Phi(\Sigma) \,.
\eqe
Another assumption is a non-increasing mass density in outward direction,
\eqb
\label{eq:ratio.normaldensity}
 \od{R(\Sigma)}{\Sigma} \le 0 \,.
\eqe
This implies the correspondence between variables $R$ and $\Sigma$ is one to one.

The surface of fluid ball is defined by vanishing pressure, where the hydrostatic equilibrium holds between fluid ball and outside vacuum region. 
Then, the mass density at surface $\Sigma_\ast$ is determined by
\eqb
\label{eq:ratio.surface}
 P(\Sigma_\ast) = 0 \,,
\eqe
and the total mass $M_\ast$ and surface radius $R_\ast$ of fluid ball are respectively given by
\eqb
 M_\ast = M(\Sigma_\ast) \quad,\quad R_\ast = R(\Sigma_\ast) \,.
\eqe
At the centre of fluid ball, the mass density is unity $\Sigma\ls{c} = 1$, and conditions, $R(1) = 0$ and $M(1) = 0$, should hold. 
Note that the mass density takes values in an interval,
\eqb
\label{eq:ratio.interval}
 \Sigma_\ast \le \Sigma \le 1 \,,
\eqe
where the surface mass density satisfies $0 \le \Sigma_\ast < 1$.

The outside region of fluid ball, $R>R_\ast$, is Schwarzschild geometry of mass $M_\ast$. 
The inside region, $R < R_\ast$, is determined by the Einstein equation and conservation law $T^{\mu\nu}_{\phantom{\mu\nu};\nu} = 0$, which are reduced to Tolman-Oppenheimer-Volkoff (TOV) equations,
\numb
\eqab
\label{eq:ratio.tov-1}
 \od{M(\Sigma)}{\Sigma} &=&
 4 \pi R(\Sigma)^2 \Sigma \od{R(\Sigma)}{\Sigma}
\\
\label{eq:ratio.tov-2}
 \od{P(\Sigma)}{\Sigma} &=& A(M,R,P;\Sigma) \od{R(\Sigma)}{\Sigma}
\\
\label{eq:ratio.tov-3}
 \od{\Phi(\Sigma)}{\Sigma} &=&
 - \frac{1}{\Sigma + P(\Sigma)} \od{P(\Sigma)}{\Sigma} \,.
\eqae
\nume
where
\eqb
\label{eq:ratio.A}
 A(M,R,P;\Sigma) \defeq
 - \frac{[\,\Sigma + P(\Sigma)\,]\,[\,M(\Sigma) + 4 \pi R(\Sigma)^3 P(\Sigma)\,]}
        {R(\Sigma)\,[\,R(\Sigma) - 2 M(\Sigma)\,]}
 \,.
\eqe
Given a concrete functional form of equation of state, two functions $R(\Sigma)$ and $M(\Sigma)$ are obtained by solving \eref{eq:ratio.tov-1} and \eref{eq:ratio.tov-2}. 
Substituting those solutions into \eref{eq:ratio.tov-3}, $\Phi(\Sigma)$ is obtained.

In addition to Buchdahl's assumptions~\eref{eq:ratio.barotropy} and~\eref{eq:ratio.normaldensity}, we assume subluminal condition of sound speed,
\eqb
\label{eq:ratio.subluminal}
 \od{P(\Sigma)}{\Sigma} \le 1 \,.
\eqe
Given the above formulation, the remaining freedom is the functional form of $P(\Sigma)$ under the condition~\eref{eq:ratio.subluminal}. 
Since the solution of TOV equations depend on the concrete form of $P(\Sigma)$, the value of total mass $M_\ast$ and surface radius $R_\ast$ of fluid ball vary with the concrete form of $P(\Sigma)$. 
In following sections, by the variation of functional form of $P(\Sigma)$ under the condition~\eref{eq:ratio.subluminal}, a possible upper bound $\bound$ of the ratio $M_\ast/R_\ast \,(= Gm_\ast/(c^2 r_\ast)\,)$ will be calculated,
\eqb
\label{eq:ratio.bound}
 \frac{M_\ast}{R_\ast} \le \bound \,.
\eqe
Using such variational method, our aim is to obtain the value of $\bound$ lower than the Barraco-Hamity's bound $\bound\ls{BaHa} = 3/8$ which is lower than the famous Buchdahl's bound $\bound\ls{Buch} = 4/9$.

Here note that Barraco and Hamity~\cite{ref:barraco+1.2002} considered not only the Buchdahl's assumptions~\eref{eq:ratio.barotropy} and~\eref{eq:ratio.normaldensity} but also the dominant energy condition which gives the following inequality for the perfect fluid ball,
\eqb
\label{eq:ratio.P-Sigma}
 P(\Sigma) \le \Sigma \,.
\eqe
We expect that some physically reasonable (non-exotic) condition, which is more restrictive than the dominant energy condition, can decrease further the Barraco-Hamity's bound to a lower value. 
Concerning this expectation, let us point out that the subluminal-sound-speed condition~\eref{eq:ratio.subluminal} and definition of surface~\eref{eq:ratio.surface} predict the inequality~\eref{eq:ratio.P-Sigma} as well. 
The subluminal-sound-speed condition restricts not only the value of pressure such as inequality~\eref{eq:ratio.P-Sigma} but also the value of its differential such as inequality~\eref{eq:ratio.subluminal}. 
This implies that the subluminal-sound-speed condition~\eref{eq:ratio.subluminal} is more restrictive than the dominant energy condition. 
Therefore, we expect that the subluminal-sound-speed condition is sufficient for us in search of an upper bound lower than Barraco-Hamity's bound.

\subsection{Variational method: Strategy}
\label{sec:ratio.strategy}

The variational method which we are going to use is known as the optimal control theory. 
It has already been applied to an estimation of maximum mass of neutron star by Rhoades and Ruffini~\cite{ref:rhoades+1.1974}. 
(See also~\cite{ref:rhoades+1.1973,ref:pietronero+1.1978} for more details about application of the optimal control theory to neutron star mass.) 
We apply the optimal control theory to the search of upper bound $\bound$ of inequality~\eref{eq:ratio.bound}. 
However, we do not require readers to have knowledge of optimal control theory. 
All ideas of the variational analysis are explained below.

We introduce an auxiliary variable $U(\Sigma)$ by
\eqb
\label{eq:ratio.U}
 \sin^2U(\Sigma) \defeq \od{P(\Sigma)}{\Sigma} \,.
\eqe
Then, the subluminal-sound-speed condition~\eref{eq:ratio.subluminal} is automatically satisfied, $\sin^2U \le 1$. 
In order to make use of variational method, it is useful to express $M_\ast/R_\ast$ by an integral form,~\footnote{
The signature of \eref{eq:ratio.M/R} is not $-\int\diff R\, \diff (M/R)/\diff R$ but $+\int\diff R\, \diff (M/R)/\diff R$, because of an inequality $\diff (M/R)/\diff R \ge 0$ which is found by $M/R \sim R^2$.}
\numb
\eqb
\label{eq:ratio.M/R}
 \frac{M_\ast}{R_\ast}
 = \int_0^{R_\ast}\!\!\!\diff R\, \od{(M/R)}{R}
 = \int_{\Sigma_\ast}^1\!\!\!\diff\Sigma\, \Bigl(- \od{(M/R)}{\Sigma} \Bigr)
 \defeqr \int_{\Sigma_\ast}^1\!\!\!\diff\Sigma\,L \,,
\eqe
where the integrand $L$ is arranged to the following form by using relation~\eref{eq:ratio.U} and TOV equations~\eref{eq:ratio.tov-1} and~\eref{eq:ratio.tov-2},
\eqb
\label{eq:ratio.L}
 L(M,R,P,U;\Sigma) =
 \frac{M - 4 \pi R^3 \Sigma}{A(M,R,P;\Sigma) R^2} \sin^2U \,.
\eqe
\nume
Our problem is to maximize the functional~\eref{eq:ratio.M/R} under three constraints; the relation \eref{eq:ratio.U} and TOV equations~\eref{eq:ratio.tov-1} and~\eref{eq:ratio.tov-2}. 
In following discussion, \eref{eq:ratio.tov-3} is not considered, because $\Phi$ does not appear in the integrand $L$.

Because there are three constraints, we use the Lagrange's multiplier-method. 
Define a functional,
\numb
\eqb
\label{eq:ratio.I}
 I \defeq
 \int_{\Sigma_\ast}^1\!\!\!\diff\Sigma
 [\, L + Y\ls{M}(\Sigma)\, C\ls{M} + Y\ls{R}(\Sigma)\, C\ls{R} + Y\ls{P}(\Sigma)\, C\ls{P} \,]
 \,,
\eqe
where functions $Y\ls{M}(\Sigma)$, $Y\ls{R}(\Sigma)$ and $Y\ls{P}(\Sigma)$ are Lagrange multiplier, and $C\ls{M}$, $C\ls{R}$ and $C\ls{P}$ are defined according to the constraints~\eref{eq:ratio.tov-1}, \eref{eq:ratio.tov-2} and~\eref{eq:ratio.U},
\eqab
 C\ls{M}(M,R,P;\Sigma)
 &=&
 \frac{4 \pi R(\Sigma)^2\Sigma)}{A(M,R,P;\Sigma)}\sin^2U(\Sigma) - \od{M(\Sigma)}{\Sigma}
\\
 C\ls{R}(M,R,P;\Sigma)
 &=&
  \frac{1}{A(M,R,P;\Sigma)}\sin^2U(\Sigma) - \od{R(\Sigma)}{\Sigma}
\\
 C\ls{P}(M,R,P;\Sigma)
 &=&
 \sin^2U(\Sigma) - \od{P(\Sigma)}{\Sigma} \,.
\eqae
\nume
Extremal value of $M_\ast/R_\ast$ is given by the condition $\delta I = 0$ under variations,
\eqb
\label{eq:ratio.variation}
\begin{array}{rcl}
 M(\Sigma) &\to& M(\Sigma) + \delta M(\Sigma)
 \\
 R(\Sigma) &\to& M(\Sigma) + \delta R(\Sigma)
 \\
 P(\Sigma) &\to& M(\Sigma) + \delta P(\Sigma)
 \\
 U(\Sigma) &\to& U(\Sigma) + \delta U(\Sigma) \,.
\end{array}
\eqe

Given the functional expression $I$ of $M_\ast/R_\ast$, the strategy of our variational analysis consists of following steps:
\begin{description}
\item[step1]
Divide the interval~\eref{eq:ratio.interval} of $\Sigma$ into many infinitesimal intervals. 
Then, solve the equation $\delta I = 0$ in order to maximize the functional $I$ at each infinitesimal interval. 
\item[step2]
Integrate the maximized $I$ of every infinitesimal interval of $\Sigma$. 
Then, we will obtain a formal expression of the upper bound $\bound \ge M_\ast/R_\ast$ for the global interval~\eref{eq:ratio.interval}. 
\item[step3]
Calculate numerically the formal expression of $\bound$ obtained in step2. 
\end{description}

\subsection{Variational method: Step1}
\label{sec:ratio.step1}

We divide the interval~\eref{eq:ratio.interval} into many infinitesimal intervals. 
Look at one infinitesimal interval of $\Sigma$,
\eqb
\label{eq:ratio.interval.variation}
 \Sigma\ls{d} \le \Sigma \le \Sigma\ls{u} \,,
\eqe
where $\Sigma_\ast \le \Sigma\ls{d} < \Sigma\ls{u} \le 1$ and $\Sigma\ls{u}-\Sigma\ls{d} \ll 1$. 
Then, in the step1 of our variational analysis, we maximize $M_\ast/R_\ast$ by solving $\delta I=0$ under the variations~\eref{eq:ratio.variation} in the interval~\eref{eq:ratio.interval.variation}. 
Note that, since we are considering a single interval~\eref{eq:ratio.interval.variation} in the step1, we fix variables ($M, R, P, U$) in the remaining intervals, $\Sigma_\ast \le \Sigma < \Sigma\ls{d}$ and $\Sigma\ls{u} < \Sigma \le 1$. 
This indicates boundary conditions,
\eqb
\label{eq:ratio.BC}
 \delta X(\Sigma\ls{d}) = 0 \quad,\quad \delta X(\Sigma\ls{u}) = 0 \quad (X=M, R, P, U) \,.
\eqe

The Euler-Lagrange equations of $\delta I = 0$ are followings:
Variation of $U$ gives
\eqb
\label{eq:ratio.ELeq.U}
 \Bigl[ \frac{M - 4 \pi R^3 \Sigma}{A R^2}
 + Y\ls{M}\frac{4 \pi R^2 \Sigma}{A} + Y\ls{R} \frac{1}{A} + Y\ls{P}
 \Bigr] \sin U \, \cos U = 0 \,,
\eqe
variations of $X (=M, R, P)$ give
\eqb
\label{eq:ratio.ELeq.X}
\fl
 -\pd{Y\ls{X}(\Sigma)}{\Sigma} =
 \Bigl[ \pd{[\,(M - 4 \pi R^3 \Sigma)/(A R^2)\,]}{X}
 + Y\ls{M} \pd{[\,4\pi R^2 \Sigma/A\,]}{X}
 + Y\ls{R} \pd{(1/A)}{X} \Bigr] \sin^2U \,,
\eqe
and variations of $Y\ls{X}$ give $C_X = 0$ which are constraints; TOV equations~\eref{eq:ratio.tov-1}, \eref{eq:ratio.tov-2}, and relation~\eref{eq:ratio.U}. 
\Eref{eq:ratio.ELeq.U} gives
\eqb
\label{eq:ratio.U.sol}
 U = 0 \,,\, \frac{\pi}{2} \,,
\eqe
or an algebraic equation given by vanishing the inside of square bracket of \eref{eq:ratio.ELeq.U}. 
However, the latter case (vanishing square bracket) is impossible under the boundary condition~\eref{eq:ratio.BC} as shown in \ref{app:U}. 
Therefore, $U$ should be constant given in \eref{eq:ratio.U.sol}, which denotes for extremal case of $M_\ast/R_\ast$,\footnote{
In the case of Buchdahl's analysis~\cite{ref:buchdahl.1959}, the maximum value $M_\ast/R_\ast = \bound\ls{Buch} = 4/9$ is realized for the fluid ball of constant mass density. 
The constant mass density implies that the sound speed is infinity. 
On the other hand, we are considering the subluminal-sound-speed condition. 
Therefore, the value $1$ of sound speed for the extremal case of $M_\ast/R_\ast$ in our analysis is consistent with Buchdahl's analysis.
}
\eqb
\label{eq:ratio.P/Sigma}
 \od{P(\Sigma)}{\Sigma} = 0 \,,\, 1 \,.
\eqe
This gives a constant pressure or linear equation of state, $P(\Sigma) = \Sigma - \Sigma_\ast$. 
It is already revealed by Nilsson and Uggla~\cite{ref:nilsson+1.2001} that, for the linear equation of state ($P=\Sigma-\Sigma_\ast$), the total mass $M_\ast$ and surface radius $R_\ast$ are both finite for $\Sigma_\ast \neq 0$, but both infinite for $\Sigma_\ast = 0$. 
However, the behaviour of $M_\ast/R_\ast$ has not been investigated so far.

\begin{figure}[t]
 \begin{center}
 \includegraphics[height=40mm]{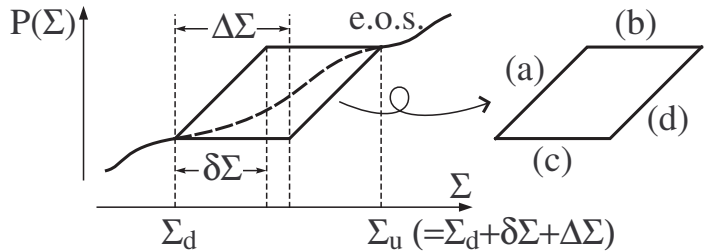}
 \end{center}
\caption{Extremization of $M_\ast/R_\ast$ in an infinitesimal interval $\Sigma\ls{d} \le \Sigma \le \Sigma\ls{u}$.
An original equation of state expressed by dashed curve is deformed to an extremal case of $M_\ast/R_\ast$, which forms an parallelogram.
}
\label{fig:parallelogram.small}
\end{figure}

Due to the extremal condition~\eref{eq:ratio.P/Sigma} and boundary condition~\eref{eq:ratio.BC}, a parallelogram appears in $\Sigma$-$P$ plane as show in \fref{fig:parallelogram.small}. 
The equation of state given by edge (a)+(b) or (c)+(d) corresponds to a maximum or minimum value of $M_\ast/R_\ast$ under the variations~\eref{eq:ratio.variation} in the infinitesimal interval~\eref{eq:ratio.interval.variation}. 
In order to judge which edge corresponds to maximum or minimum, we calculate their difference,
\numb
\eqb
 \Delta\Bigl(\frac{M_\ast}{R_\ast}\Bigr)
 \defeq \int\ls{(c)+(d)}\!\!\!\diff\Sigma\, L - \int\ls{(a)+(b)}\!\!\!\diff\Sigma\, L \,,
\eqe
where the first term corresponds to an integral with equation of state given by the edge (c)+(d), and the second term is an integral with the edge (a)+(b). 
Here, note that TOV equations~\eref{eq:ratio.tov-1} and~\eref{eq:ratio.tov-2} result in that $M(\Sigma)$ and $R(\Sigma)$ are constant for $U = 0$ ($\diff P/\diff \Sigma = 0$). 
Therefore, the edges (b) and (c) cause no contribution to $M_\ast/R_\ast$. 
Then, the difference $\Delta(M_\ast/R_\ast)$ is calculated to be
\eqb
\label{eq:ratio.DeltaM/R}
\fl
\eqalign{
 \Delta\Bigl(\frac{M_\ast}{R_\ast}\Bigr)
 &=
 \int\ls{(d)}\!\!\!\diff\Sigma\, L - \int\ls{(a)}\!\!\!\diff\Sigma\, L
\\
 &=
 L(M,R,P,U=1;\Sigma\ls{d}+\Delta\Sigma)\,\delta\Sigma - L(M,R,P,U=1;\Sigma\ls{d})\,\delta\Sigma
\\
 &=
 \pd{L(M,R,P,U=1;\Sigma\ls{d})}{\Sigma\ls{d}}
    \delta\Sigma\,\Delta\Sigma
\\
 &=
 \frac{R\ls{d}-2 M\ls{d}}{R\ls{d} (\Sigma\ls{d}+P\ls{d})^2}\delta\Sigma\,\Delta\Sigma
 > 0 \,,
}
\eqe
\nume
where $\delta\Sigma$ and $\Delta\Sigma$ correspond to two edges of parallelogram as shown in \fref{fig:parallelogram.small}, and $M\ls{d}$, $R\ls{d}$ and $P\ls{d}$ are mass, radius and pressure at $\Sigma\ls{d}$. 
This inequality~\eref{eq:ratio.DeltaM/R} denotes that the equation of state with edge (c)+(d) (edge (a)+(b)\,) corresponds to the maximum (minimum) of $M_\ast/R_\ast$ under variations in the infinitesimal interval~\eref{eq:ratio.interval.variation} and boundary condition~\eref{eq:ratio.BC}.

\subsection{Variational method: Step2}
\label{sec:ratio.step2}

\begin{figure}[t]
 \begin{center}
 \includegraphics[height=45mm]{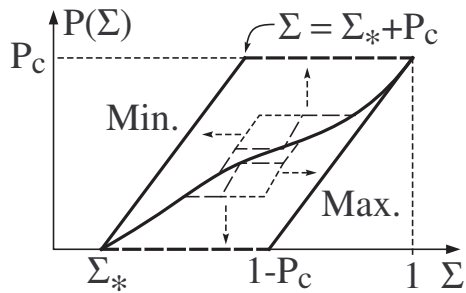}
 \end{center}
\caption{Maximization and minimization of $M_\ast/R_\ast$ in the full interval $\Sigma_\ast \le \Sigma \le 1$ with fixing $P\ls{c}$ and $\Sigma_\ast$. 
Horizontal dashed parts of e.o.s. do not contribute to $M_\ast/R_\ast$.
}
\label{fig:parallelogram.large}
\end{figure}

Next, we proceed to an analysis in the global interval~\eref{eq:ratio.interval} of $\Sigma$. 
Consider a case that the values of pressure at centre, $P\ls{c}$, and mass density at surface, $\Sigma_\ast$, are fixed. 
In this case, we can repeat the deformation of equation of state using an infinitesimally small parallelogram. 
Finally, the functional form of equation of state in interval~\eref{eq:ratio.interval} reaches one of following two options:
\numb
\eqab
\label{eq:ratio.eos.minimum}
 P(\Sigma)
 &=&
 \Biggl\{
 \begin{array}{lcl}
 \Sigma - \Sigma_\ast & \mbox{in} & \Sigma_\ast \le \Sigma \le \Sigma_\ast + P\ls{c}
 \\
 P\ls{c} & \mbox{in} & \Sigma_\ast + P\ls{c} < \Sigma \le 1
 \end{array}
\\
\label{eq:ratio.eos.maximum}
 P(\Sigma)
 &=&
 \Biggl\{
 \begin{array}{lcl}
 0 & \mbox{in} & \Sigma_\ast \le \Sigma < 1-P\ls{c}
 \\
 \Sigma - (1-P\ls{c}) & \mbox{in} & 1-P\ls{c} \le \Sigma \le 1
 \end{array} \,.
\eqae
\nume
These forms are shown in \fref{fig:parallelogram.large}. 
No contribution to $M_\ast/R_\ast$ causes from parts, $P=$ constant, as explained before calculating \eref{eq:ratio.DeltaM/R}. 
The inequality~\eref{eq:ratio.DeltaM/R} for infinitesimally small parallelogram indicates that the equation of state~\eref{eq:ratio.eos.maximum} gives the maximum value of $M_\ast/R_\ast$ for given $P\ls{c}$ and $\Sigma_\ast$, whereas \eref{eq:ratio.eos.minimum} gives the minimum for given $P\ls{c}$ and $\Sigma_\ast$.

We should emphasize that the above discussion is applicable for only the case with fixed $P\ls{c}$ and $\Sigma_\ast$. 
It has not been examined so far whether or not $M_\ast/R_\ast$ for equation of state~\eref{eq:ratio.eos.maximum} (or \eref{eq:ratio.eos.minimum}) is maximum (or minimum) even when the values of $P\ls{c}$ and $\Sigma_\ast$ vary. 
There is a possibility that a value of $M_\ast/R_\ast$ for equations of state~\eref{eq:ratio.eos.minimum} or~\eref{eq:ratio.eos.maximum} for certain values of $P\ls{c}$ and $\Sigma_\ast$ is neither maximum nor minimum under variation of $P\ls{c}$ and $\Sigma_\ast$.

In order to find a true maximum value of $M_\ast/R_\ast$ under variation of $P\ls{c}$ and $\Sigma_\ast$, we regard $M_\ast/R_\ast$ for equation of state~\eref{eq:ratio.eos.maximum} as a function of $P\ls{c}$, 
\eqb
\label{eq:ratio.f}
 f(P\ls{c}) \defeq
 \frac{M_\ast}{R_\ast}\,\, \mbox{for equation of state~\eref{eq:ratio.eos.maximum}} \,.
\eqe
Here, note that the equation of state~\eref{eq:ratio.eos.maximum} in the interval, $1-P\ls{c} \le \Sigma \le 1$, depends on only $P\ls{c}$. 
The maximum value of $f(P\ls{c})$, which is denoted by $\bound$, is the desired upper bound of mass-to-radius ratio, $M_\ast/R_\ast \le \bound$. 
A numerical plot of $f(P\ls{c})$ is going to be shown in next section~\ref{sec:ratio.step3}.

In order to find a true minimum value of $M_\ast/R_\ast$ under variation of $P\ls{c}$ and $\Sigma_\ast$, we regard $M_\ast/R_\ast$ for equation of state~\eref{eq:ratio.eos.minimum} as a function of $P\ls{c}$ and $\Sigma_\ast$. 
The minimum value of this function is the lower bound of $M_\ast/R_\ast$. 
Here, note that this lower bound can be read from $f(P\ls{c})$ by following discussion: 
For the equation of state~\eref{eq:ratio.eos.minimum} for given $P\ls{c}$ and $\Sigma_\ast$, the quantity $\Sigma_\ast + P\ls{c}$ can be regarded as a central mass density.\footnote{
We should also remember, $R(\Sigma) =$ constant for $P =$ constant due to TOV equations.}
Then, following the normalization~\eref{eq:ratio.normalization}, we transform variables by $\widetilde{\Sigma} = \Sigma/(\Sigma_\ast + P\ls{c})$, $\widetilde{P} = P/(\Sigma_\ast + P\ls{c})$, $\widetilde{M} = \sqrt{\Sigma_\ast + P\ls{c}}\, M$ and $\widetilde{R} = \sqrt{\Sigma_\ast + P\ls{c}}\, R$. 
This transformation changes the form of equation of state~\eref{eq:ratio.eos.minimum} in interval, $\Sigma_\ast \le \Sigma \le \Sigma_\ast + P\ls{c}$, to the following form,
\eqb
\label{eq:ratio.eos.transform}
 \widetilde{P}(\widetilde{\Sigma}) = \widetilde{\Sigma} - (1-\widetilde{P}\ls{c})
 \quad\mbox{for}\quad
 1-\widetilde{P}\ls{c} \le \widetilde{\Sigma} \le 1 \,,
\eqe
where $\widetilde{P}\ls{c} \defeq P\ls{c}/(\Sigma_\ast + P\ls{c})$. 
Furthermore, the form of TOV equations for transformed variables are the same with \eref{eq:ratio.tov-1} and~\eref{eq:ratio.tov-2}. 
Hence, the value of $\widetilde{M}_\ast/\widetilde{R}_\ast$ for equation of state~\eref{eq:ratio.eos.transform} is equal to the value of $M_\ast/R_\ast$ for equation of state~\eref{eq:ratio.eos.minimum}. 
On the other hand, the form of equation of state~\eref{eq:ratio.eos.transform} is the same with equation of state~\eref{eq:ratio.eos.maximum} in interval, $1-P\ls{c} \le \Sigma \le 1$. 
Therefore, the value of $M_\ast/R_\ast$ for equation of state~\eref{eq:ratio.eos.minimum} for given $P\ls{c}$ and $\Sigma_\ast$ is equal to that for equation of state~\eref{eq:ratio.eos.maximum} with replacing $P\ls{c}$ by $\widetilde{P}\ls{c}$. 
This implies that the value of $M_\ast/R_\ast$ for equation of state~\eref{eq:ratio.eos.minimum} can be read from $f(P\ls{c})$, and the minimum value of $f(P\ls{c})$ is the lower bound of $M_\ast/R_\ast$.

\subsection{Variational method: Step3}
\label{sec:ratio.step3}

\begin{figure}[t]
 \begin{center}
 \includegraphics[height=80mm]{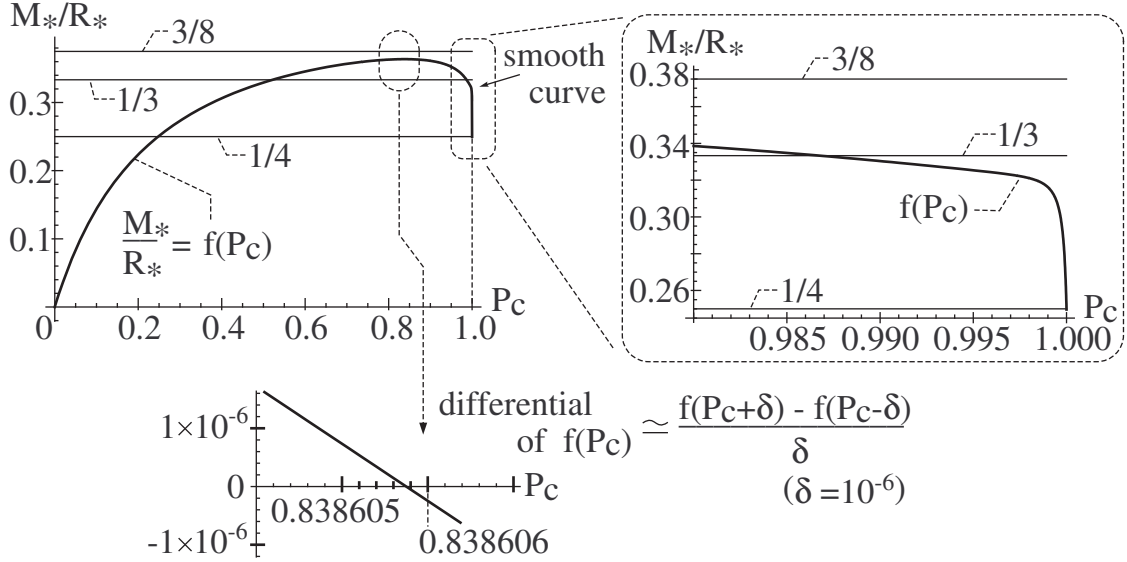}
 \end{center}
\caption{$M_\ast/R_\ast$ as a function of $P\ls{c}$ for linear equation of state~\eref{eq:ratio.eos.maximum}. 
Differential is plotted by an approximation, $\diff f(P\ls{c})/\diff P\ls{c} \simeq [\,f(P\ls{c}+\delta)-f(P\ls{c}-\delta)\,]/\delta$, where $\delta = 10^{-6}$.
These plots are made with Mathematica ver.10.}
\label{fig:result}
\end{figure}

The function $f(P\ls{c})$, defined in \eref{eq:ratio.f}, can be calculated by following steps: 
\begin{description}
\item[(i)]
Solve numerically TOV equations~\eref{eq:ratio.tov-1} and~\eref{eq:ratio.tov-2} with the equation of state~\eref{eq:ratio.eos.maximum} for given value of $P\ls{c}$. 
A technical remark is summarized in \ref{app:tov}. 
\item[(ii)]
Calculate $M_\ast/R_\ast$ from numerical solutions of step (i). 
\item[(iii)] The function $f(P\ls{c})$ is calculated by repeating steps (i) and (ii) for different values of $P\ls{c}$ in the interval, $0 < P\ls{c} \le 1$. 
\end{description}
Our numerical result is shown in \fref{fig:result}. 
A smooth graph of $f(P\ls{c})$ is obtained. 
From the plot of differential of $f(P\ls{c})$ in \fref{fig:result}, we find that the maximum of $f(P\ls{c})$ is given at
\eqb
\label{eq:ratio.Pbound}
 P\ls{c} = P\ls{bound} \simeq 0.8386058 \,. 
\eqe
Then, as discussed in previous section~\ref{sec:ratio.step2}, the upper bound $\bound\ls{new} \ge M_\ast/R_\ast$ is given by the maximum value, $\bound\ls{new} \simeq f(P\ls{bound})$, which is evaluated numerically to be $\bound\ls{new} \simeq 0.3636403$. 
This upper bound gives our conclusion~\eref{eq:intro.result}. 
On the other hand, the lower bound read from \fref{fig:result} is zero given at $P\ls{c} = 0$. 
No finite lower bound is found in our analysis.

\section{Conclusion and discussions}
\label{sec:cd}

By adding the subluminal-sound-speed condition~\eref{eq:ratio.subluminal} to Buchdahl's assumptions, we obtained \fref{fig:result} and the bound of mass-to-radius ratio of a barotropic fluid ball,
\eqb
\label{eq:cd.result}
 0 < \frac{M_\ast}{R_\ast} \le \bound\ls{new} \simeq 0.3636403 \,.
\eqe
This upper bound is lower than Barraco-Hamity's bound $\bound\ls{BaHa} = 3/8$. 
However, our upper bound is greater than $1/3\, (< \bound\ls{new})$, and there remains a possibility of an existence of super-compact object possessing unstable circular orbits of photons but no black hole horizon.

As read from \fref{fig:result}, the upper bound $\bound\ls{new}$ corresponds to the central pressure, $P\ls{c} = P\ls{bound}$ of~\eref{eq:ratio.Pbound}. 
This value of central pressure may be understood as a result of trade-off between pressure's two effects; a contribution to mass density (attractive force) and an effect pushing fluid outward (repulsive force).

From the above, the maximum value of $M_\ast/R_\ast$ is $\bound\ls{new}$ of~\eref{eq:cd.result} which is obtained by solving the TOV equations with the central pressure $P\ls{c} = P\ls{bound}$ of~\eref{eq:ratio.Pbound} and linear equation of state~\eref{eq:ratio.eos.maximum}. 
Furthermore, by our numerical integration of TOV equations, the mass and radius of fluid ball in the case of maximum mass-to-radius ratio are
\eqb
\label{eq:cd.MRbound}
\begin{array}{rll}
 M_\ast(P\ls{bound}) &= M\ls{bound} &\simeq 0.2014578
\\
 R_\ast(P\ls{bound}) &= R\ls{bound} &\simeq 0.5540028 \,,
\end{array}
\eqe
which gives our bound $\bound\ls{new} = M\ls{bound}/R\ls{bound}$. 
From these values together with the normalization~\eref{eq:ratio.normalization}, we can estmate typical \emph{un-normalized} values of physical quantitis for the case of maximum mass-to-radius ratio. 
With denoting the un-normalized mass in unit of solar mass, $m\ls{bound} = \alpha M_\odot$, where $\alpha$ is a dimension-less factor, then the un-normalized central mass density $\sigma\ls{bound}$, central pressure $p\ls{bound}$ and radius $r\ls{bound}$ are calculated as
\eqb
\begin{array}{rll}
 \sigma\ls{bound}
 &\displaystyle = \frac{c^6 M\ls{bound}^2}{G^3 m\ls{bound}^2}
 &\displaystyle \simeq \frac{2.5}{\alpha^2} \times 10^{16} \,\,\,\mbox{g/cm$^3$}
\\
 p\ls{bound}
 &= \sigma\ls{bound} c^2 P\ls{bound}
 &\displaystyle \simeq \frac{1.3}{\alpha^2} \times 10^{39} \,\,\,\mbox{GeV/cm$^3$}
\\
 r\ls{bound}
 &\displaystyle = \frac{G m\ls{bound}}{c^2 \bound\ls{new}}
 &\simeq 4.1\, \alpha \,\,\,\mbox{km} \,.
\end{array}
\eqe

Note that, the case of maximum mass-to-radius ratio corresponds to the linear equation of state~\eref{eq:ratio.eos.maximum} giving the luminal (equal-to-light) sound speed inside the fluid ball. 
If the equation of state, $P(\Sigma)$, will be restricted to some functional form other than the linear form, then the upper bound $\bound$ of $M_\ast/R_\ast$ will take a value different from $\bound\ls{new}$ of~\eref{eq:cd.result}. 
We will report the case of polytropic equation of state in the other paper.\footnote{
We will show in the other paper that, when the equation of state is restricted to be the polytropic one ($P(\Sigma) \propto \Sigma^{1+1/n}$) under the subluminal-sound-speed condition, the maximum value of $M_\ast/R_\ast$ becomes a value lower than $\bound\ls{new}$ of~\eref{eq:cd.result}, $M_\ast/R_\ast \le 0.281$. 
In deriving this result of polytrope, we will not use the variational method, but use the other numerical search method.
}

Finally we make a mathematical comment on a limiting behaviour found from \fref{fig:result},
\eqb
\label{eq:conc.limit}
 \lim_{P\ls{c}\to 1}f(P\ls{c}) = \frac{1}{4} \,.
\eqe
At the limit $P\ls{c} \to 1$, the equation of state~\eref{eq:ratio.eos.maximum} asymptotes to a simple form $P(\Sigma) = \Sigma$ $(0 \le \Sigma \le 1)$. 
Further, functional forms of $M(\Sigma)$ and $R(\Sigma)$ converge to an exact solution of TOV equations with equation of state $P(\Sigma) = \Sigma$\,;
\eqb
 M(\Sigma) = \frac{1}{4} R(\Sigma) = \frac{1}{16 \sqrt{\pi \Sigma}} \,.
\eqe
This results in a finite limit~\eref{eq:conc.limit}. 
However, the total mass and surface radius of fluid ball diverge $M_\ast \to \infty$ and $R_\ast \to \infty$ as $P\ls{c} \to 1$, because the surface mass density is zero, $\Sigma_\ast \to 0$ as $P\ls{c} \to 1$. 
This result is consistent with Nilsson-Uggla's numerical result~\cite{ref:nilsson+1.2001} that, for the linear equation of state, $P = \Sigma - \Sigma_\ast$ ($\Sigma_\ast = 1 - P\ls{c}$), the mass $M_\ast$ and radius $R_\ast$ are both finite for $\Sigma_\ast \neq 0$ but diverge as $\Sigma_\ast \to 0$.

\ack

We would like to express our gratitude to Tomohiro Harada, since a communication with him lets us to notice the use of variational method. 
H.S. was supported by Japan Society for the Promotion of Science (JSPS), Grant-in-Aid for Scientific Research (KAKENHI, Exploratory Research, 26610050).

\appendix
\section{Impossibility of vanishing square bracket of \eref{eq:ratio.ELeq.U}}
\label{app:U}

In this appendix, we show that, as a solution of \eref{eq:ratio.ELeq.U}, the vanishing square bracket of \eref{eq:ratio.ELeq.U} is impossible under the boundary condition~\eref{eq:ratio.BC}.

Assume that a solution of \eref{eq:ratio.ELeq.U} is given by vanishing the square bracket, we find
\eqb
\label{eq:U.ELeq-U}
 Y\ls{P} = -\frac{1}{A(M,R,P;\Sigma)}
 \Biggl[ Y\ls{R} + \frac{M}{R^2}
 + 4 \pi R^2\,\Sigma\,\Bigl( Y\ls{M} - \frac{1}{R} \Bigr)\,\Biggr] \,.
\eqe
On the other hand, we obtain from \eref{eq:ratio.ELeq.X},
\eqab
\label{eq:U.ELeq-M}
 \od{Y\ls{M}}{\Sigma}
 &=&
 - \frac{R+8 \pi R^3 P}{(R-2M) (M+4 \pi R^3 P)}Y\ls{P}\,\sin^2U
 - \frac{\sin^2U}{A(M,R,P;\Sigma) R^2}
\\
\nonumber
 \od{Y\ls{R}}{\Sigma}
 &=&
 \Biggl( \frac{1}{R} + \frac{1}{(R-2M)} - \frac{12 \pi R^2 P}{M+4 \pi R^3 P} \Biggr)
 Y\ls{P} \sin^2U
\\
\label{eq:U.ELeq-R}
 &&
 + \Biggl[ \frac{2M}{R^3} + 4 \pi \Sigma \Bigl(1 - 2 R Y\ls{M} \Bigr)\,\Biggr]
   \frac{\sin^2U}{A(M,R,P;\Sigma)}
\\
\label{eq:U.ELeq-P}
 \od{Y\ls{P}}{\Sigma}
 &=&
 - \frac{M \Sigma + 4 \pi R^3 (P \Sigma + \Sigma + P)}{(\Sigma + P) (M + 4 \pi R^3 P)}
   Y\ls{P} \sin^2U \,.
\eqae
\Eref{eq:U.ELeq-P} with boundary condition~\eref{eq:ratio.BC} results in $Y\ls{P} = 0$. 
Then \eref{eq:U.ELeq-U} gives
\eqb
\label{eq:U.ELeq-U.2}
 Y\ls{R} + \frac{M}{R^2} + 4 \pi R^2\,\Sigma\,\Bigl( Y\ls{M} - \frac{1}{R} \Bigr) = 0 \,.
\eqe
Differentiate this equation by $\Sigma$ and substitute \eref{eq:U.ELeq-M}, \eref{eq:U.ELeq-R} and TOV equations~\eref{eq:ratio.tov-1} and~\eref{eq:ratio.tov-2}, we obtain
\eqb
 Y\ls{M}(\Sigma) = \frac{1}{R(\Sigma)} \quad,\quad
 Y\ls{R}(\Sigma) = - \frac{M(\Sigma)}{R(\Sigma)^2} \,,
\eqe
where $Y\ls{R}$ is obtained by substituting $Y\ls{M}$ in \eref{eq:U.ELeq-U.2}. 
These solutions cannot satisfy the boundary condition~\eref{eq:ratio.BC}. 
Hence, the solution of \eref{eq:ratio.ELeq.U} is given by \eref{eq:ratio.U.sol}.

\section{On numerical treatment of TOV equations}
\label{app:tov}

Right-hand sides of TOV equations~\eref{eq:ratio.tov-1} and~\eref{eq:ratio.tov-2} are indeterminate form at centre because of the conditions, $M \to 0$ and $R \to 0$ as $\Sigma \to 1$. 
Therefore, in solving TOV equations numerically in step (i) of section~\ref{sec:ratio.step3}, we have made use of perturbative solutions near the centre.

In order to consider a perturbation near centre, we regard the radius $R$ as an independent variable, and the mass density as a function of radius, $\Sigma(R)$. 
TOV equations~\eref{eq:ratio.tov-1} and~\eref{eq:ratio.tov-2} are rearranged to the following forms,
\eqb
\label{eq:tov.tov}
\fl
 \od{M(R)}{R} = 4 \pi R^2 \Sigma(R)
 \quad,\quad
 \od{P(R)}{R} =
 - \frac{\bigl[\,\Sigma(R) + P(R)\,\bigr]\,\bigl[\,M(R) + 4 \pi R^3 P(R)\,\bigr]}
        {R \,\bigl[\,R - 2 M(R)\,\bigr]} \,.
\eqe
For a sufficiently small radius $R \ll 1$, we introduce perturbations,
\eqb
\label{eq:tov.perturbation.primitive}
\eqalign{
 M(R) &= M\ls{(1)} R + M\ls{(2)} R^2 + M\ls{(3)} R^3 + \cdots
\\
 P(R) &= P\ls{c} + P\ls{(1)} R + P\ls{(2)} R^2 + P\ls{(3)} R^3 + \cdots
\\
 \Sigma(R) &= 1 + \Sigma\ls{(1)} R + \Sigma\ls{(2)} R^2 + \Sigma\ls{(3)} R^3 + \cdots \,,
}
\eqe
where conditions $M(R=0) = 0$, $\Sigma(R=0) = 1$ and $P(R=0) = P\ls{c}$ are included. 
Substituting~\eref{eq:tov.perturbation.primitive} into \eref{eq:tov.tov}, we obtain $M\ls{(1)} = 0$, $M\ls{2} = 0$, $P\ls{(1)} = 0$ and remaining parts,
\eqb
\label{eq:tov.perturbation}
\eqalign{
 M(R) &= \frac{4}{3}\pi R^3 + \pi \Sigma\ls{(1)} R^4 + \cdots
\\
 P(R)
 &= P\ls{c} - \frac{2}{3}\pi (1 + 3 P\ls{c})\,(1 + P\ls{c}) R^2
   - \frac{\pi}{9} (7 + 15 P\ls{c}) \Sigma\ls{(1)} R^3 + \cdots
\\
 \Sigma(R) &= 1 + \Sigma\ls{(1)} R + \Sigma\ls{(2)} R^2 + \Sigma\ls{(3)} R^3 + \cdots \,,
}
\eqe
where the central pressure $P\ls{c}$ and coefficients $\Sigma_{(n)}$ $(n = 1, 2, 3, \cdots)$ are determined by the concrete form of equation of state.

Comparing $P(R)$ in \eref{eq:tov.perturbation} with equations of state~\eref{eq:ratio.eos.maximum}, we obtain $\Sigma\ls{(1)} = 0$ and $\Sigma\ls{(2)} = P\ls{(2)}$. 
Hence, denoting a small radius by $R_\delta \ll 1$, the mass density $\Sigma_\delta$ and mass $M_\delta$ at $R=R_\delta$ are approximately given by
\eqb
 \Sigma_\delta = 1 - \frac{2}{3}\pi (1 + 3 P\ls{c})\,(1 + P\ls{c}) R_\delta^2
 \quad,\quad
 M_\delta = \frac{4}{3}\pi R_\delta^3 \,.
\eqe
If the mass density near centre $\Sigma_\delta$ is given, the others are determined by
\eqb
\label{eq:tov.IC}
\fl
 R_\delta = \sqrt{\frac{3 (1 - \Sigma_\delta)}{2 \pi (1+3P\ls{c}) (1+P\ls{c})}}
 \quad,\quad
 M_\delta =
 \frac{4}{3}\pi \Biggl(\frac{3 (1 - \Sigma_\delta)}{2 \pi (1+3P\ls{c}) (1+P\ls{c})}\Biggr)^{3/2}
 \,.
\eqe
In step (i) of section~\ref{sec:ratio.step3}, we have solved TOV equations~\eref{eq:ratio.tov-1} and~\eref{eq:ratio.tov-2} in an interval, $\Sigma_\ast \le \Sigma \le \Sigma_\delta$, with initial condition~\eref{eq:tov.IC}. 
Also, the convergence of numerical solutions has been checked by varying $\Sigma_\delta$. 
Our results in \fref{fig:result} are made using $\Sigma_\delta = 1 - 10^{-6}$.



\end{document}